\documentclass[a4paper]{jpconf}
\usepackage{graphicx}

\newcommand{\PT}{\ensuremath{p_{\rm T}}{}}

\newcommand{\pPbEnergy}{\ensuremath{\sqrt{s_{\mbox{\tiny NN}}}=5.02~\mathrm{TeV}}{}}
\newcommand{\ppEnergyHigh}{\ensuremath{\sqrt{s}=7~\mathrm{TeV}}{}}

\newcommand{\RAA}{\ensuremath{R_{\rm{AA}}}{}}
\newcommand{\RpA}{\ensuremath{R_{\rm{pA}}~}{}}

\newcommand{\qTPC}{\ensuremath{q_{2}^{\rm TPC}~}{}}

\begin{document}
\title{Open heavy flavour production in heavy--ion
collisions with ALICE}

\author{Martin V\"olkl for the ALICE Collaboration}

\address{Eberhard Karls Universit\"at T\"ubingen, Geschwister-Scholl-Platz, 72074 T\"ubingen, Germany}

\ead{martin.andreas.volkl@cern.ch}

\begin{abstract}
Heavy--flavour measurements give an important contribution to the understanding of the Quark--Gluon Plasma (QGP) produced in heavy--ion collisions. Heavy quarks are effective probes for the QGP and understanding their interaction with the medium can give insights into its transport properties. The ALICE Collaboration investigates heavy--flavour production by measuring leptons from heavy--flavour decays and by reconstructing hadronic decays of D mesons and charmed baryons. In this proceeding, recent results from ALICE as well as the direction of future heavy--flavour measurements are discussed.
\end{abstract}

\section{Introduction}
Heavy quarks -- charm and beauty -- are important tools to study the properties of the hot and dense matter produced in relativistic heavy--ion collisions. The study of their interaction with the medium can yield insight into its properties. Due to their large mass ($m_{\rm c/b}\approx 1.5/4.5~{\rm GeV}/c$), heavy quarks are produced almost exclusively in the initial hard scattering processes of the collision. Therefore, they witness the full evolution of the collision. Additionally, the large quark masses make perturbative calculations applicable even to a production at rest.
In the vast majority of models, the interaction is modeled as discrete scatterings of the heavy quark off the medium constituents. In this framework, collisional ($2 \rightarrow 2$) and radiative ($2 \rightarrow 3$) processes can be distinguished. The relative contribution of these processes differs between models. It depends on the momentum and mass of the heavy quark. As a consequence, measurements over a large momentum range for both flavours are needed to disentangle the contributions.

After the initial hard scatterings, the number of heavy quarks in the system stays constant throughout the interaction. Thus, the effect of the medium is a redistribution of the produced particles in momentum space. Fast heavy quarks are slowed down, while slow ones may be accelerated due to the collective flow of the medium around them. The nuclear modification factor is frequently used to quantify the effects due to the formation of a medium. It compares the effect of a sample of heavy--ion collisions with an estimated mean value of binary (nucleon--nucleon) collisions of $\langle N_{\rm coll} \rangle$ to a superposition of the same number of proton--proton (pp) collisions:

\begin{equation}
R_\mathrm{AA}(p_{\mathrm{T}})=\frac{1}{\langle N_{\mathrm{coll}}\rangle} \frac{\mathrm{d} N_\mathrm{AA}/\mathrm{d} p_\mathrm{T}}{\mathrm{d} N_\mathrm{pp}/\mathrm{d} p_\mathrm{T}}=\frac{1}{\langle T_{\mathrm{AA}}\rangle} \frac{\mathrm{d} N_\mathrm{AA}/\mathrm{d} p_\mathrm{T}}{\mathrm{d} \sigma_\mathrm{pp}/\mathrm{d} p_\mathrm{T}} ~,
\end{equation}
where $T_{\mathrm{AA}}$ is the nuclear overlap function. This approach assumes that the individual nucleon--nucleon collisions are well represented by proton--proton collisions in the vacuum. In reality, the fact that the nucleons are part of a large nucleus may give rise to cold--nuclear--matter (CNM) effects which have to be estimated to interpret the measured nuclear modification factor. Additional measurements are performed in p--Pb collisions to disentangle CNM effects from those due to the medium.
Another important observable to quantify the interaction of heavy quarks with the medium is the elliptic flow coefficient $v_2$. It is defined as $v_2 = \langle \cos[2(\phi-\Psi_2)] \rangle$, the second Fourier coefficient with respect to the second order symmetry plane of the azimuthal distribution. The $v_2$ quantifies the degree of participation in the collective motion of the surrounding medium.

\begin{figure}[h]
\begin{minipage}{0.45\textwidth}
\includegraphics[height=17pc]{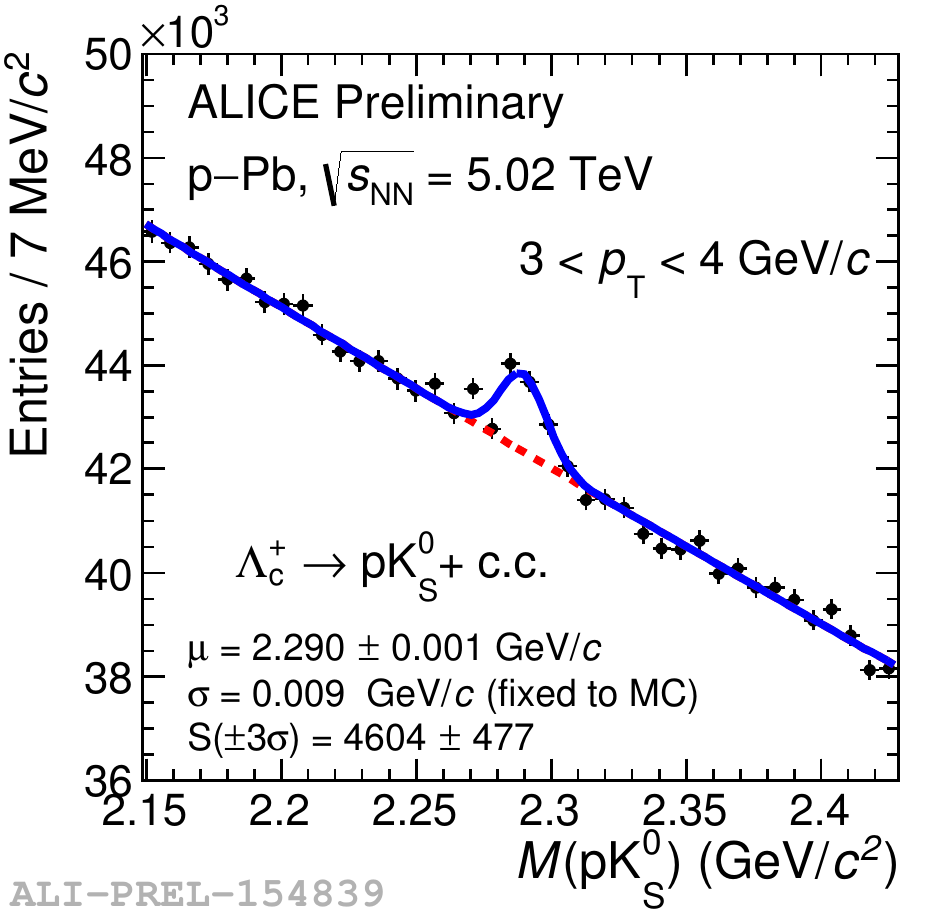}
\caption{\label{invMass}Example of an invariant mass peak in the fully hadronic reconstruction of $\Lambda_c^{+}$.}
\end{minipage}\hspace{0.1\textwidth}%
\begin{minipage}{0.45\textwidth}
\includegraphics[height=17pc]{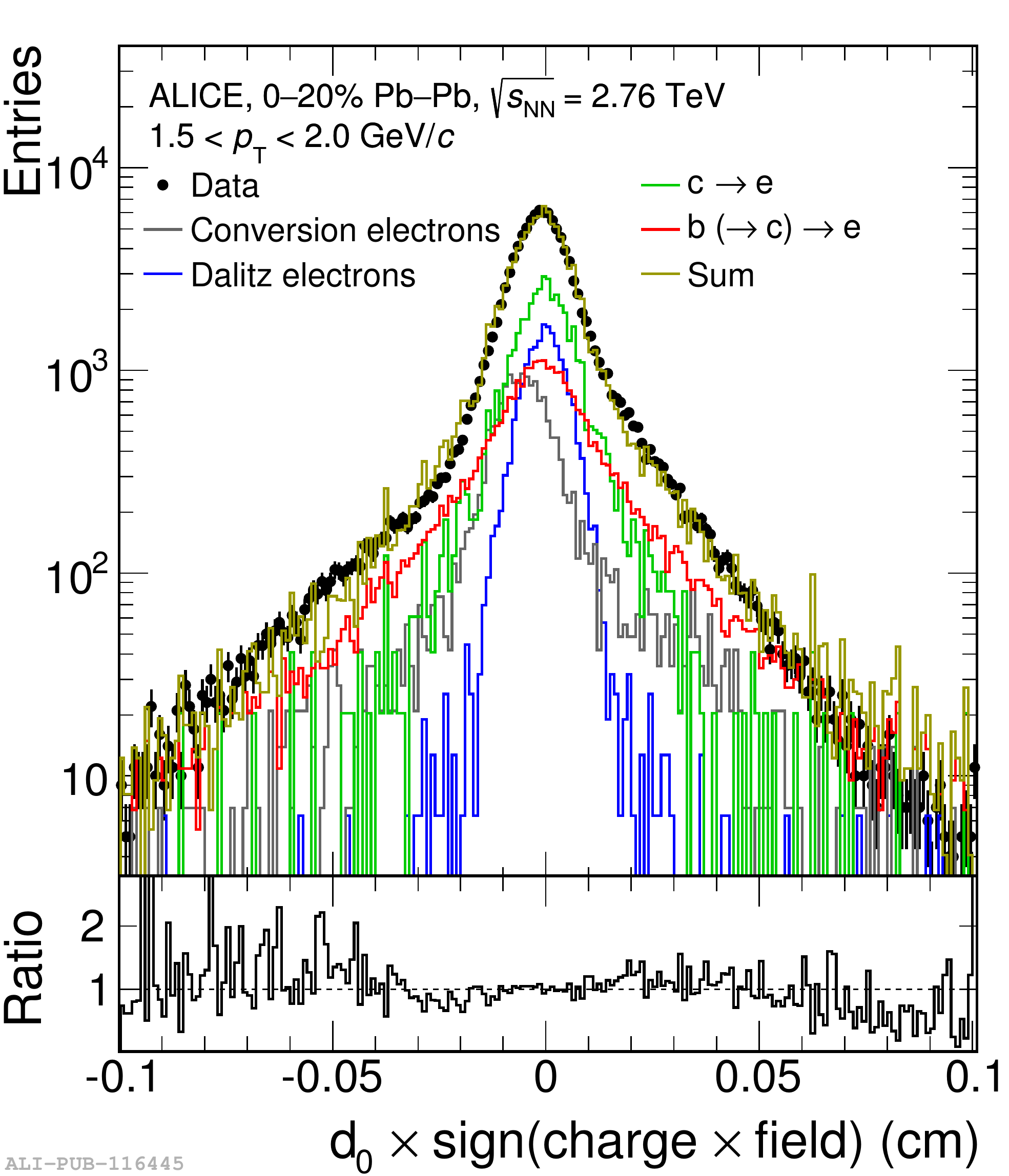}
\caption{\label{DCAFit}Example of the transverse impact parameter template fit for extracting the beauty contribution to the electrons \cite{Adam:2016wyz}.}
\end{minipage} 
\end{figure}

\section{Experimental methods}
The measurement at low transverse momenta and particle identification are particular strengths of the ALICE experiment. At mid--rapidity, open charm hadrons are measured via their hadronic decays (an example is shown in Figure \ref{invMass}). This includes $D^0 \rightarrow K^{-} \pi^{+}$ (branching ratio $3.93 \pm 0.04 \%$), $D^{+} \rightarrow K^{-} \pi^{+}\pi^{+}$ ($9.46 \pm 0.24 \%$), $D^{*+} \rightarrow D^{0} \pi^{+}$ ($67.7 \pm 0.5 \%$), $D_{s}^{+} \rightarrow \phi^{-} \pi^{+}$ ($2.27 \pm 0.08 \%$). The $\Lambda^{+}_{c}$ is reconstructed via the $p K^{-} \pi^{+}$ and $p K^{0}$ channels and via the semi--leptonic $e^{-} \nu_{e} \Lambda$ channel. The $\Xi^{0}_{c}$ is reconstructed via the semi--leptonic $e^{+}\Xi^{-}\nu_e$ channel. The decay products can be identified using the signals of the Time Projection Chamber (TPC) and the Time Of Flight detector (TOF). For measurements at forward rapidity and to gain an access to beauty, semileptonic decays of heavy flavour hadrons are used. These are decays of the type $b \rightarrow l +X$, where only the lepton $l$ is measured. Such a measurement has the advantage that just by using particle identification, a large amount of the background can be removed. Due to the fairly large branching ratios and the softer \PT--distributions of electrons from background sources, electrons with transverse momenta above a few ${\rm GeV}/c$ mostly come from heavy flavour sources. The branching ratios of charmed hadrons to a final state containing electrons is about $10\%$. For beauty hadrons, the ratio is about $20\%$, with about half of the contribution coming from decays with intermediate charmed states. At forward rapidity ($2.5<\eta<4$ for symmetric collisions), heavy flavour muons are measured by the Muon Spectrometer. The central barrel measurements ($|\eta|<0.8$) are based on electrons identified using the TPC and TOF with additional PID information from the Inner Tracking System (ITS) at low \PT~and the Electromagnetic Calorimeter (EMCal) at high \PT. To extract the contribution of heavy flavours, an estimate for the light--flavour contribution is subtracted from the measured leptons. To separate the beauty contribution, the impact parameter of the electron in the transverse plane is used as additional information. The different electron sources are then separated by a template fit of the transverse impact parameter distributions (example in Figure \ref{DCAFit}).

\begin{figure}[h]
\begin{minipage}{0.30\textwidth}
\vspace{0.5pc}
\includegraphics[height=13pc]{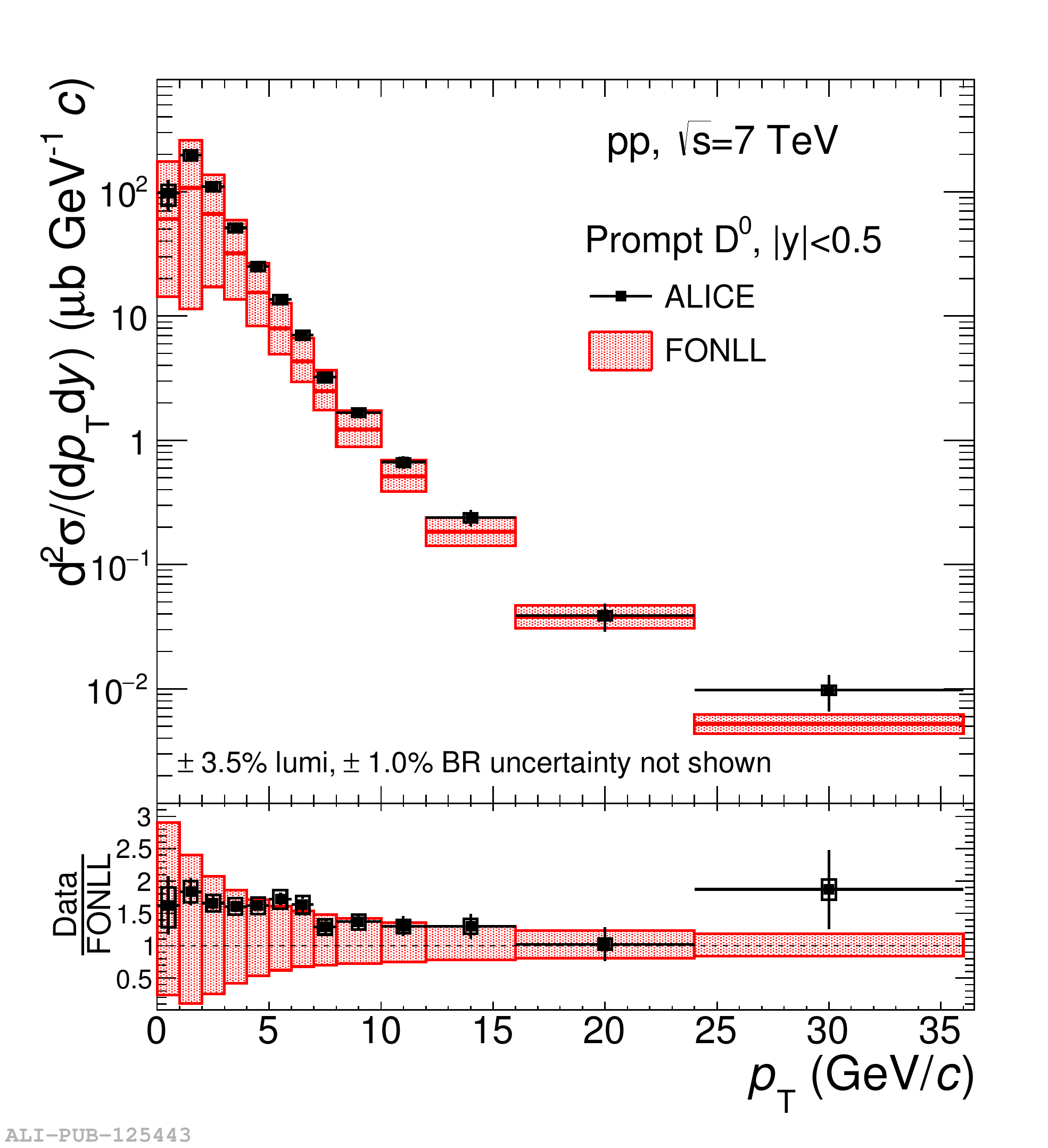}
\vspace{0.5pc} 
\caption{\label{ppDMesons}$D^{0}$ cross section measurement as function of transverse momentum \cite{Acharya:2017jgo}.}
\end{minipage}\hspace{0.05\textwidth}%
\begin{minipage}{0.30\textwidth}
\includegraphics[height=15pc]{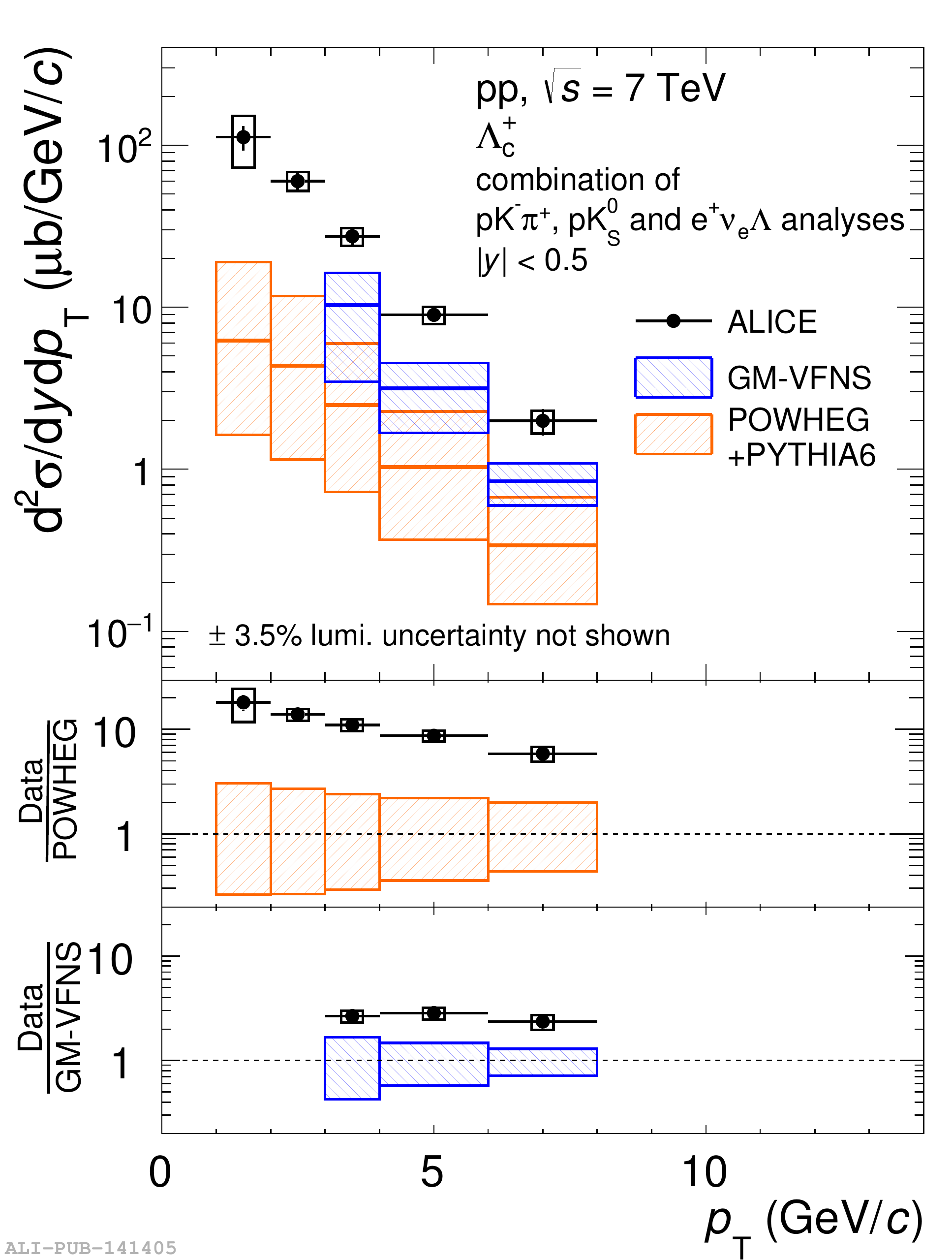}
\caption{\label{Lambdacpp}$\Lambda_c^{+}$ measurement in pp collisions at \ppEnergyHigh \cite{Acharya:2017kfy}.}
\end{minipage}\hspace{0.05\textwidth}%
\begin{minipage}{0.30\textwidth}
\includegraphics[height=15pc]{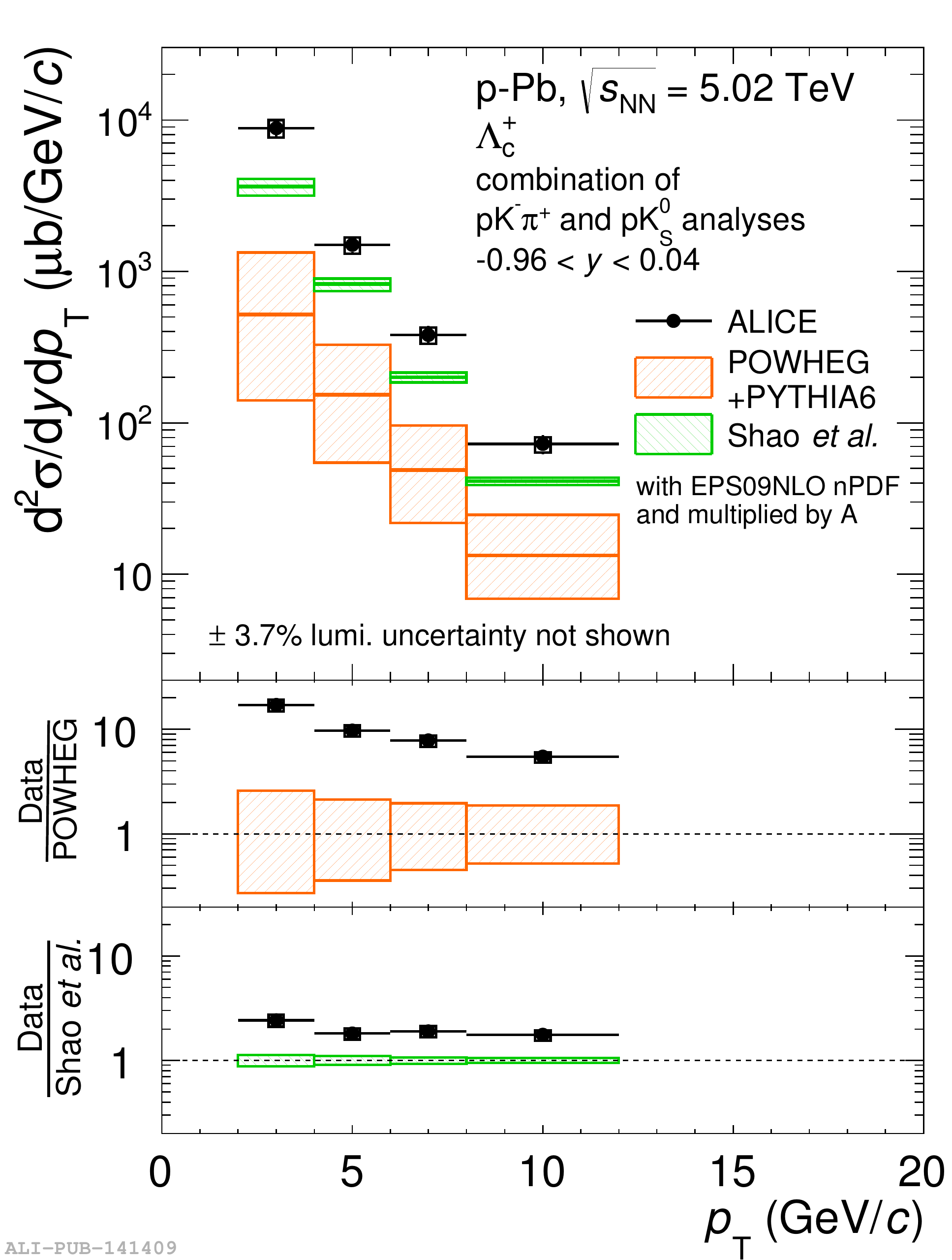}
\caption{\label{LambdacpPb}$\Lambda_c^{+}$ measurement in p--Pb collisions at \pPbEnergy \cite{Acharya:2017kfy}.}
\end{minipage}  
\end{figure}

\section{Results in small systems} 
Measurements in pp give the needed baseline to calculate the nuclear modification factor \RAA. In addition they can test the quality of our theoretical understanding of the production processes. Figure \ref{ppDMesons} shows a measurement of the production cross sections of $D^{0}$ mesons in pp collisions at \ppEnergyHigh. The measurement uncertainties are much smaller than those from the theoretical calculations. The central points are near the upper edge of the calculations. Another important physical process is the hadronization. It is inherently a nonperturbative process, thus the experimental input is particularly important for its understanding. Additionally, it is of interest to study further not just how heavy flavours are distributed in phase space but also how they are distributed among the possible hadron species. In particular, the charmed meson to baryon ratio can give insight into the charm hadronization mechanism. Figures \ref{Lambdacpp} and \ref{LambdacpPb} show the measured yield of $\Lambda_c^{+}$ in pp and p--Pb collisions together with model calculations. The models underestimate the production of $\Lambda_c^{+}$ by about a factor of 10 (for the central values) in both collision systems \cite{Acharya:2017kfy}. This also provides an important baseline to compare the ratios in Pb--Pb with. In addition, the yield of $\Xi_c^{0}$ was measured. Even considering the large uncertainties in the branching ratio, the measured yields are much larger than suggested by the models.

\begin{figure}[h]
\begin{minipage}{0.45\textwidth}
\includegraphics[height=15pc]{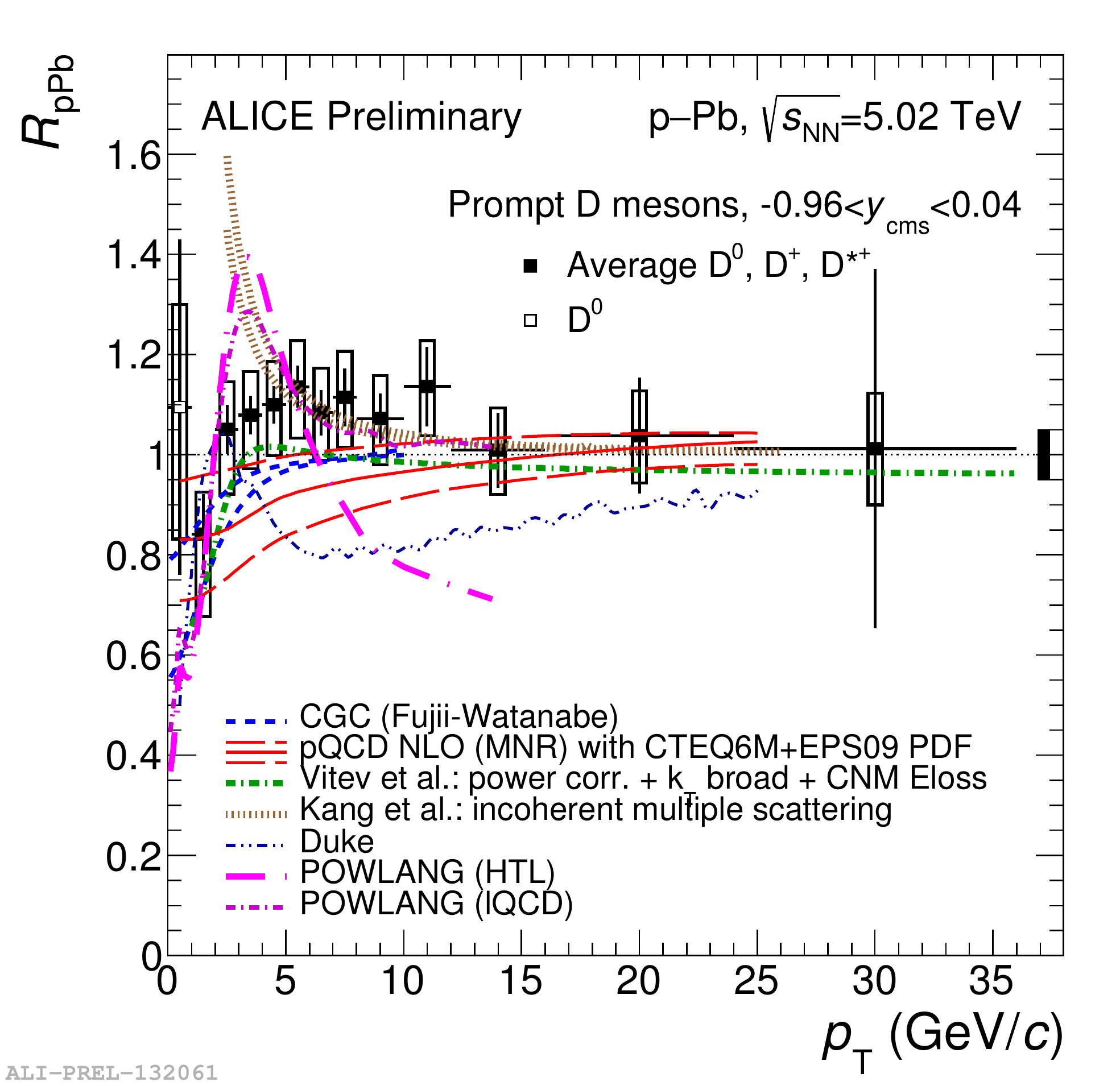}
\caption{\label{DRpA}\RpA of D mesons in p--Pb collisions at \pPbEnergy.}
\end{minipage}\hspace{0.05\textwidth}%
\begin{minipage}{0.5\textwidth}
\vspace{1.5pc}
\includegraphics[height=13pc]{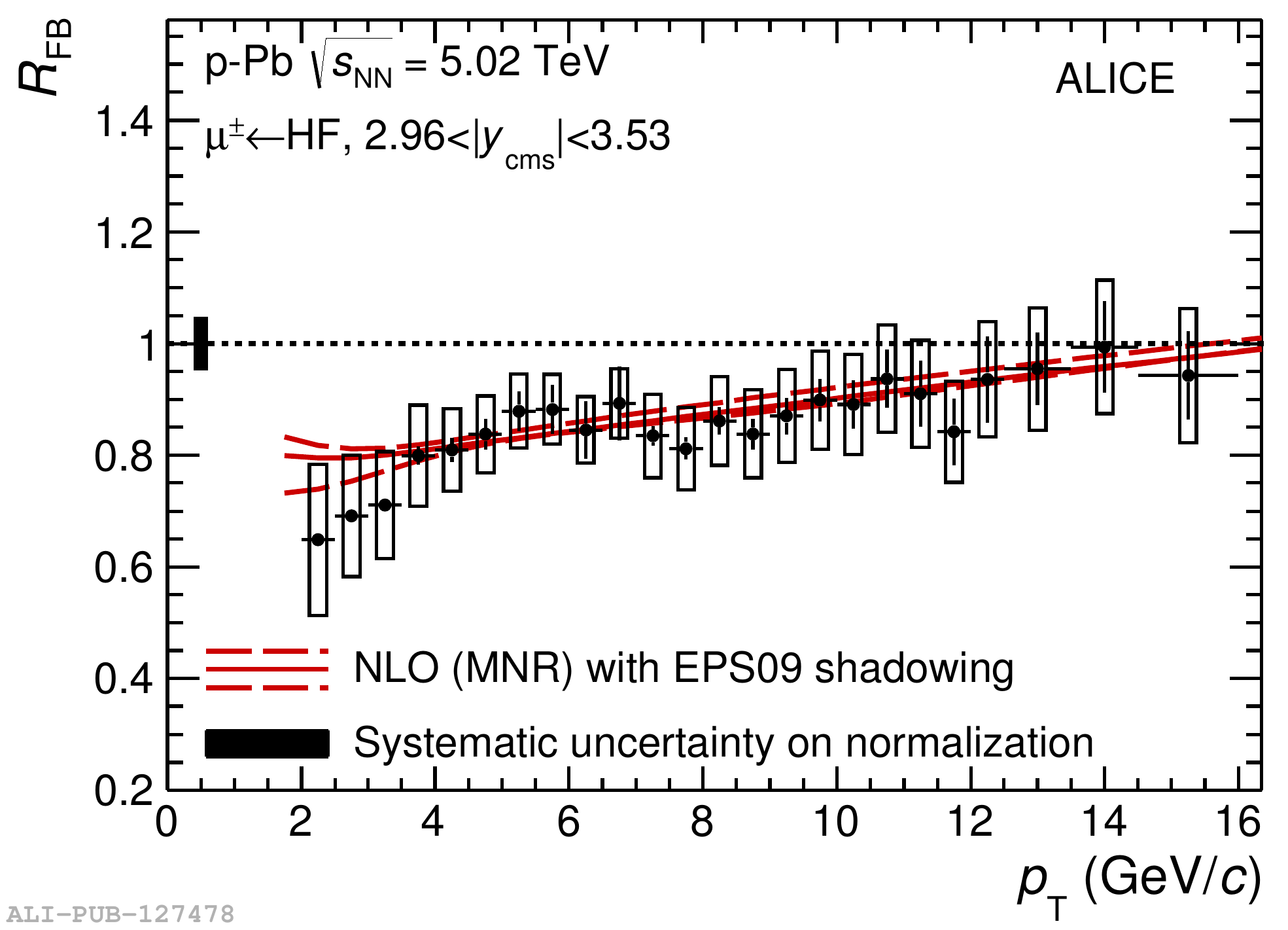}
\vspace{1.5pc}
\caption{\label{MuonRFB}Ratio of \RpA of muons from heavy flavour hadron decays measured in p--Pb and Pb--p collisions}
\end{minipage} 
\end{figure}

Comparison of the yields in pp and p--Pb collisions can aid in the understanding of cold nuclear matter effects. While the results for the \RpA of reconstructed D mesons shown in Figure \ref{DRpA} broadly follow the trend of the calculations based on nuclear parton distribution functions (nPDFs), no strong deviation from unity is found either. To gain a more thorough understanding of CNM effects it has proven useful to compare measurements at forward and backward rapidity. This is possible by using the Muon Spectrometer in p--Pb collisions and Pb--p collisions, which can be achieved by switching the beams. Both in the model calculations and in the measurements, the ratio of the forward and backward nuclear modification factors $R_{\rm FB}$ has small uncertainties, because many contributions cancel out in the ratio. The $R_{\rm FB}$ of muon from heavy--flavour hadron decays is shown in Figure \ref{MuonRFB}. It deviates from unity and is consistent with the expectations from NLO calculations which include nPDFs \cite{Acharya:2017hdv}.

\begin{figure}[h]
\begin{minipage}{0.45\textwidth}
\includegraphics[height=15pc]{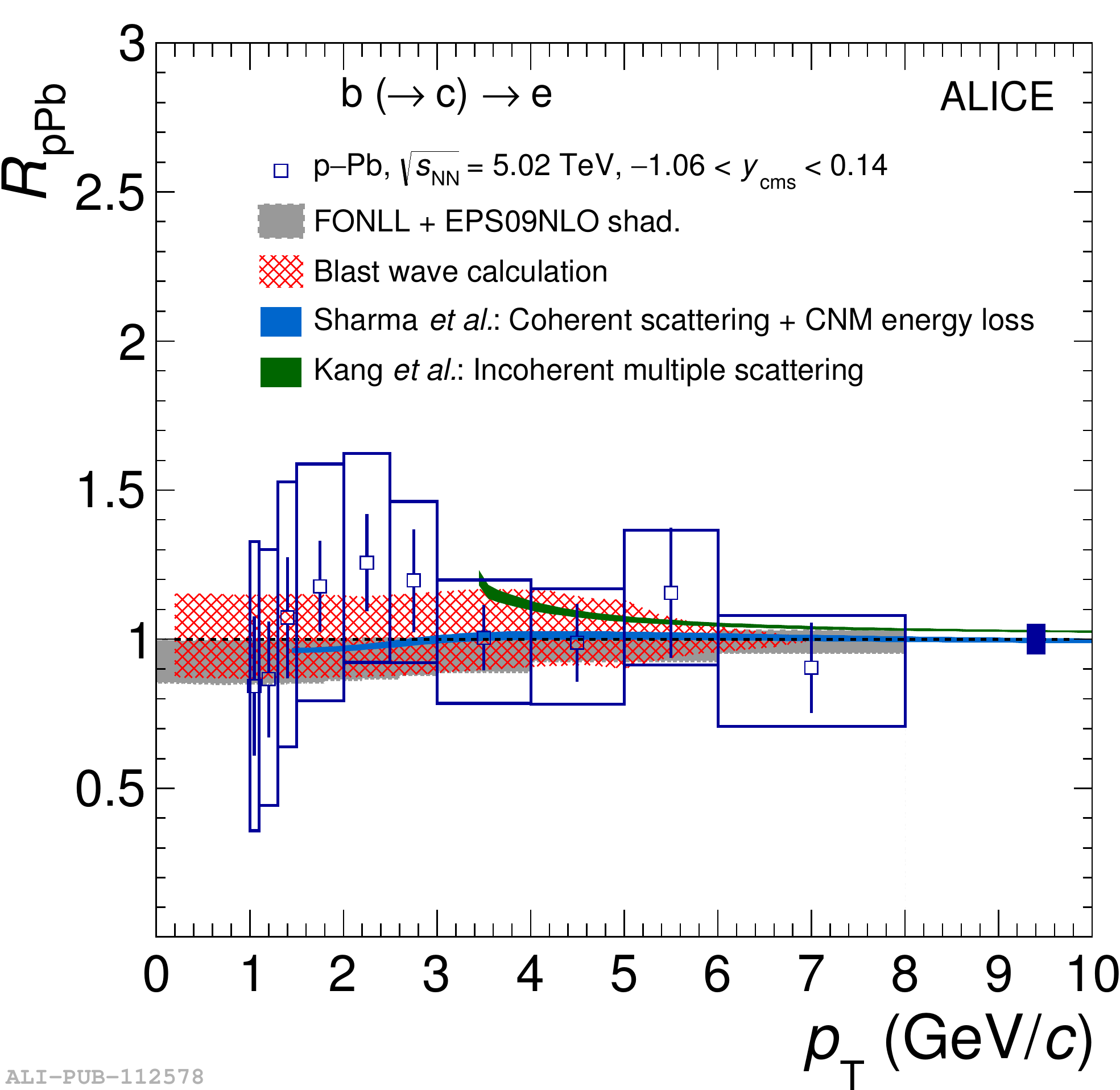}
\caption{\label{BtoeRpA}\RpA of electrons from beauty--hadron decays \cite{Adam:2016wyz}.}
\end{minipage}\hspace{0.05\textwidth}%
\begin{minipage}{0.5\textwidth}
\vspace{1.pc}
\includegraphics[height=13pc]{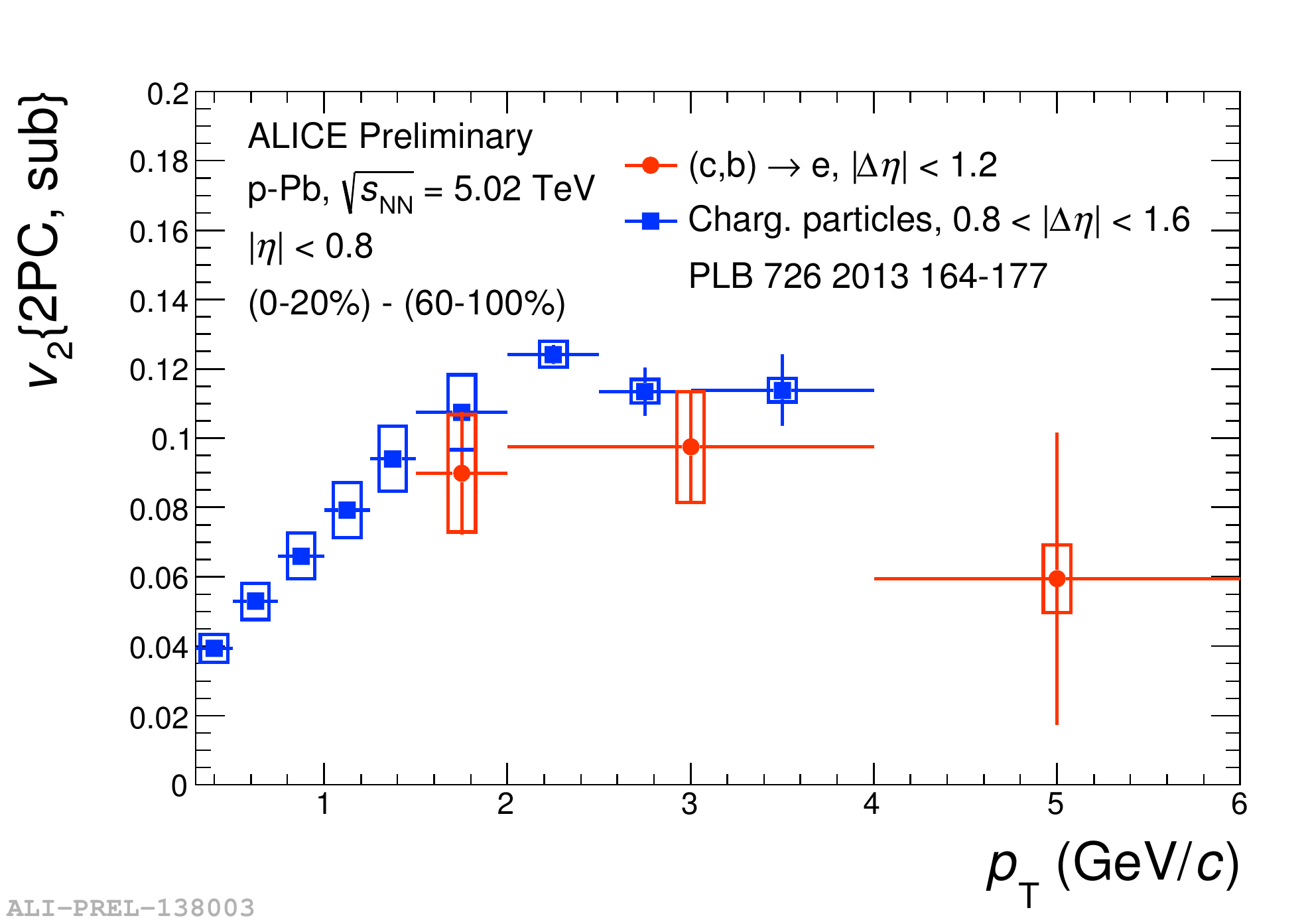}
\vspace{1.pc}
\caption{\label{HFeFlowpPb}Elliptic flow of electrons from heavy--flavour hadron decays in p--Pb collisions.}
\end{minipage} 
\end{figure}

Measuring electrons from beauty--hadron decays specifically, also yields an \RpA consistent with unity as shown in Figure \ref{BtoeRpA}. All these results point towards pp and p--Pb collisions not showing large modifications of the heavy flavour production. As a consequence, strong deviations from the model of vacuum superpositions of nucleon--nucleon collisions in Pb--Pb could be interpreted as being medium induced. However, measurements of collective flow in small systems challenge this simple picture. Figure \ref{HFeFlowpPb} shows a measured elliptic flow coefficient for electrons form heavy--flavour decays, which is significantly larger than zero. It is similar in maginitude to the measurement of light particles \cite{ABELEV:2013wsa}. The measurement of collective behaviour in the angular distributions, simultaneous with essentially unmodified \PT--distributions requires careful additional studies.

\begin{figure}[h]
\begin{minipage}{0.45\textwidth}
\includegraphics[height=15pc]{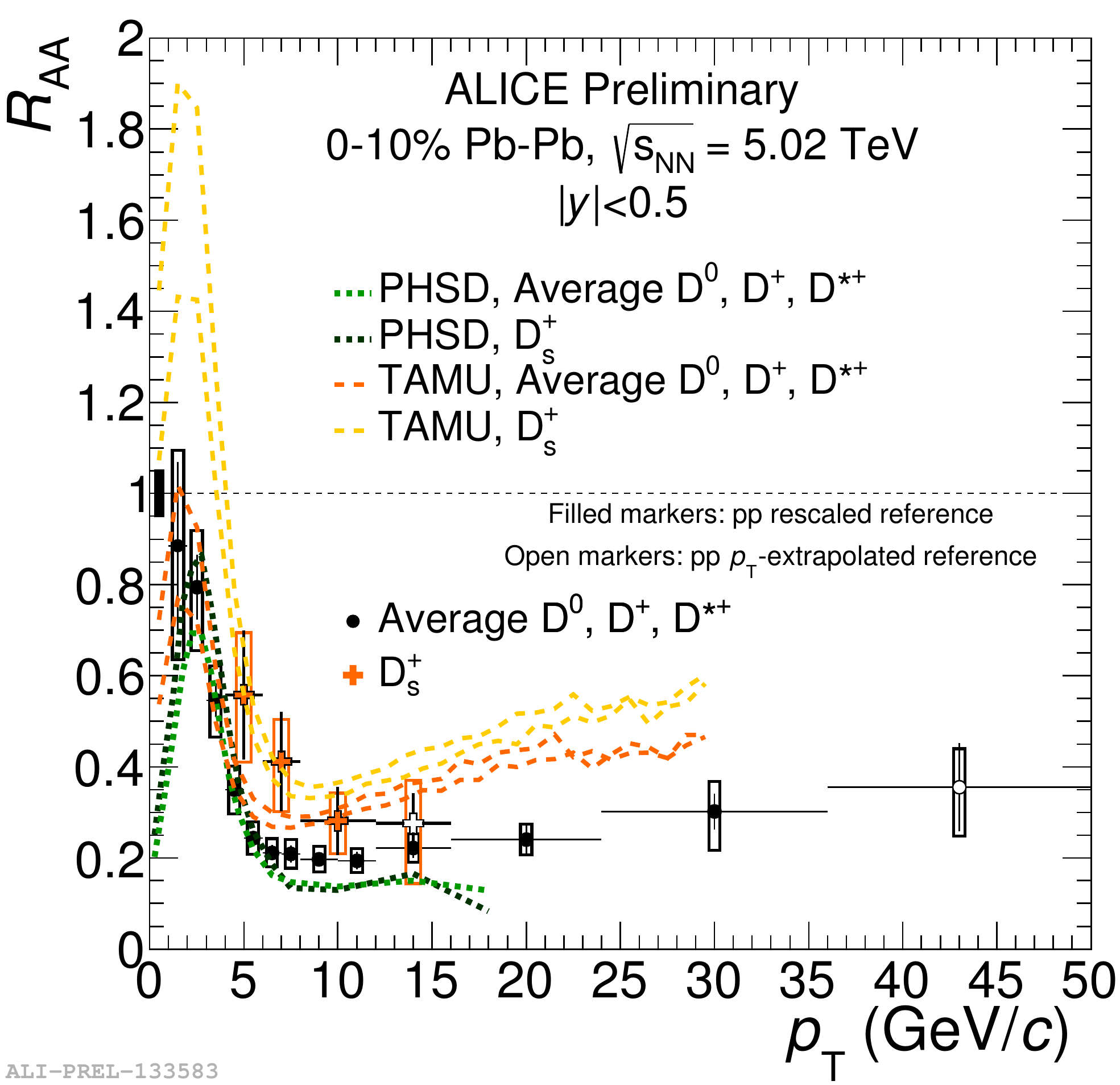}
\caption{\label{DRAA}\RAA of D mesons in central Pb--Pb collisions.}
\end{minipage}\hspace{0.1\textwidth}%
\begin{minipage}{0.45\textwidth}
\vspace{2.pc}
\includegraphics[height=11pc]{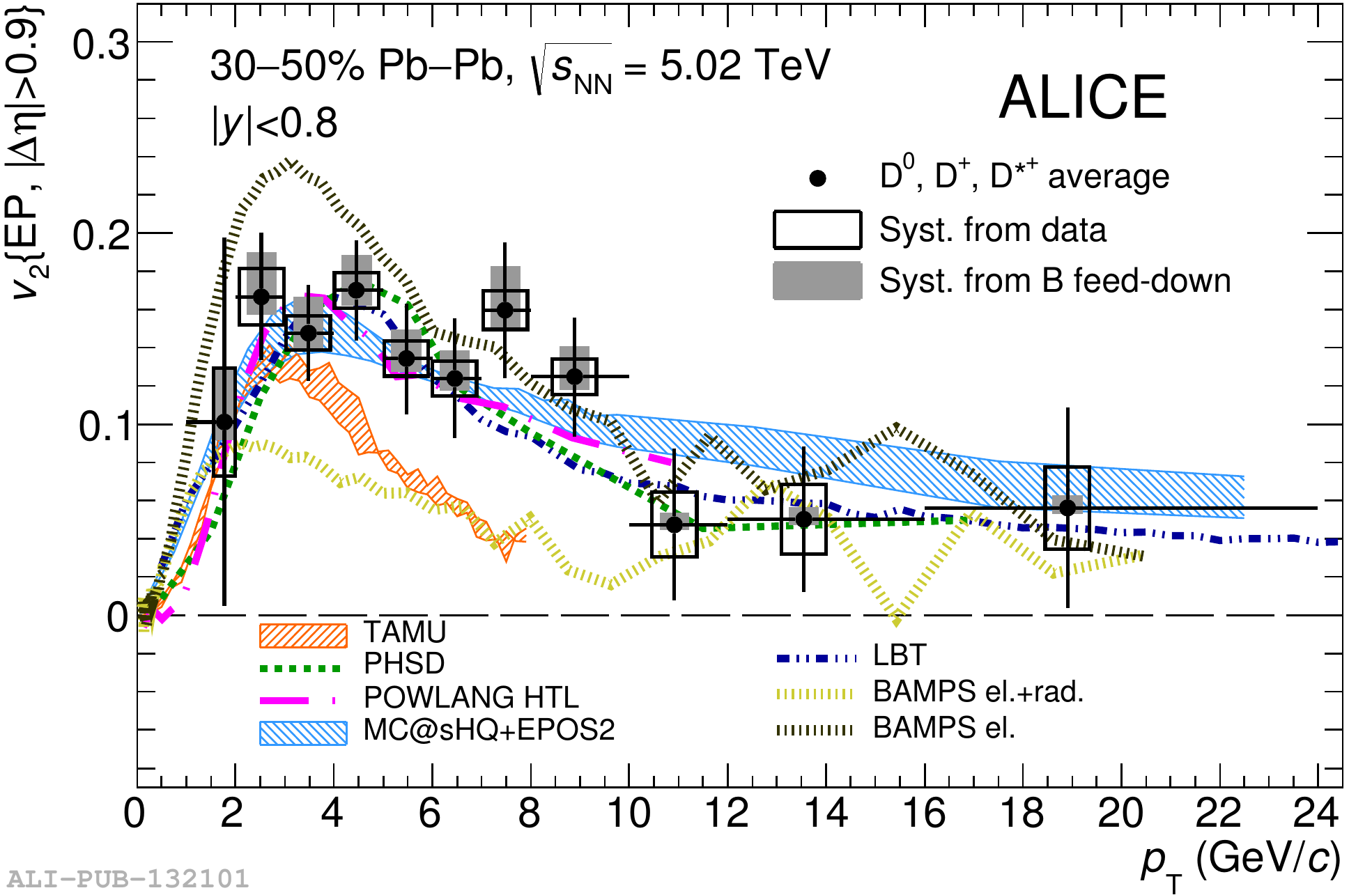}
\vspace{2.pc}
\caption{\label{Dv2}Elliptic flow of D mesons in Pb--Pb collisions.}
\end{minipage} 
\end{figure}

\section{Results in Pb--Pb collisions}

The \PT--distributions of D mesons show a strong modification in central Pb--Pb collisions as shown in Figure \ref{DRAA}. In addition, the elliptic flow of D mesons (Figure \ref{Dv2}) shows a magnitude similar to that of light hadrons. This suggests a thermalization time for charm quarks of the order of the system lifetime. Careful comparison of the most successful models suggests $\tau_c = 3-14 ~{\rm fm}/c$. The nuclear modification factor of $D_s^{+}$ was also measured. The result at low to intermediate \PT~ hints at a larger \RAA with respect to non--strange D mesons at the same \PT, consistent with an increased production of strange mesons expected by model predictions.

\begin{figure}[h]
\begin{minipage}{0.45\textwidth}
\includegraphics[height=17pc]{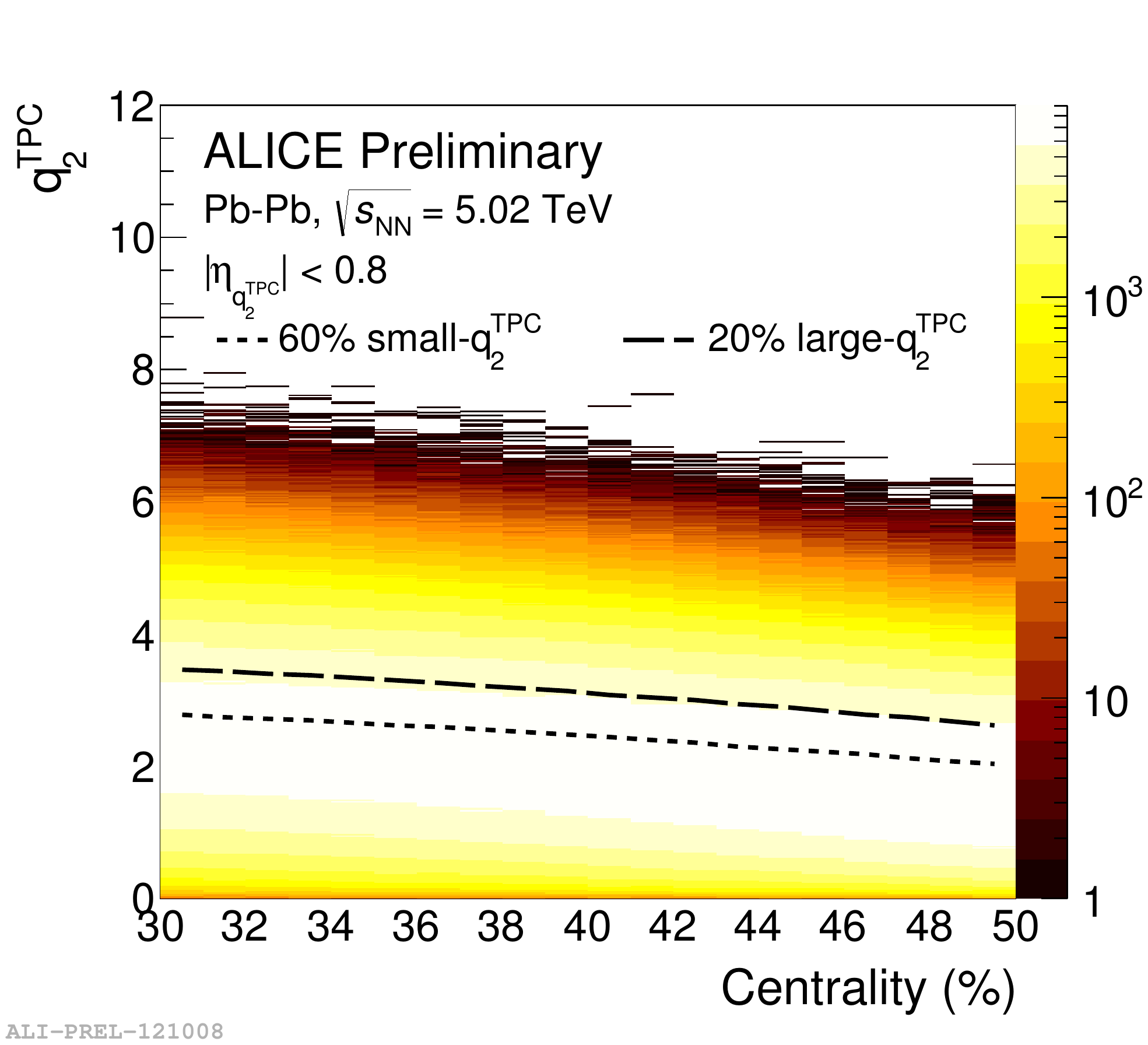}
\caption{\label{qVector}Distribution of reduced second--order flow vector \qTPC for different centralities.}
\end{minipage}\hspace{0.1\textwidth}%
\begin{minipage}{0.45\textwidth}
\includegraphics[height=17pc]{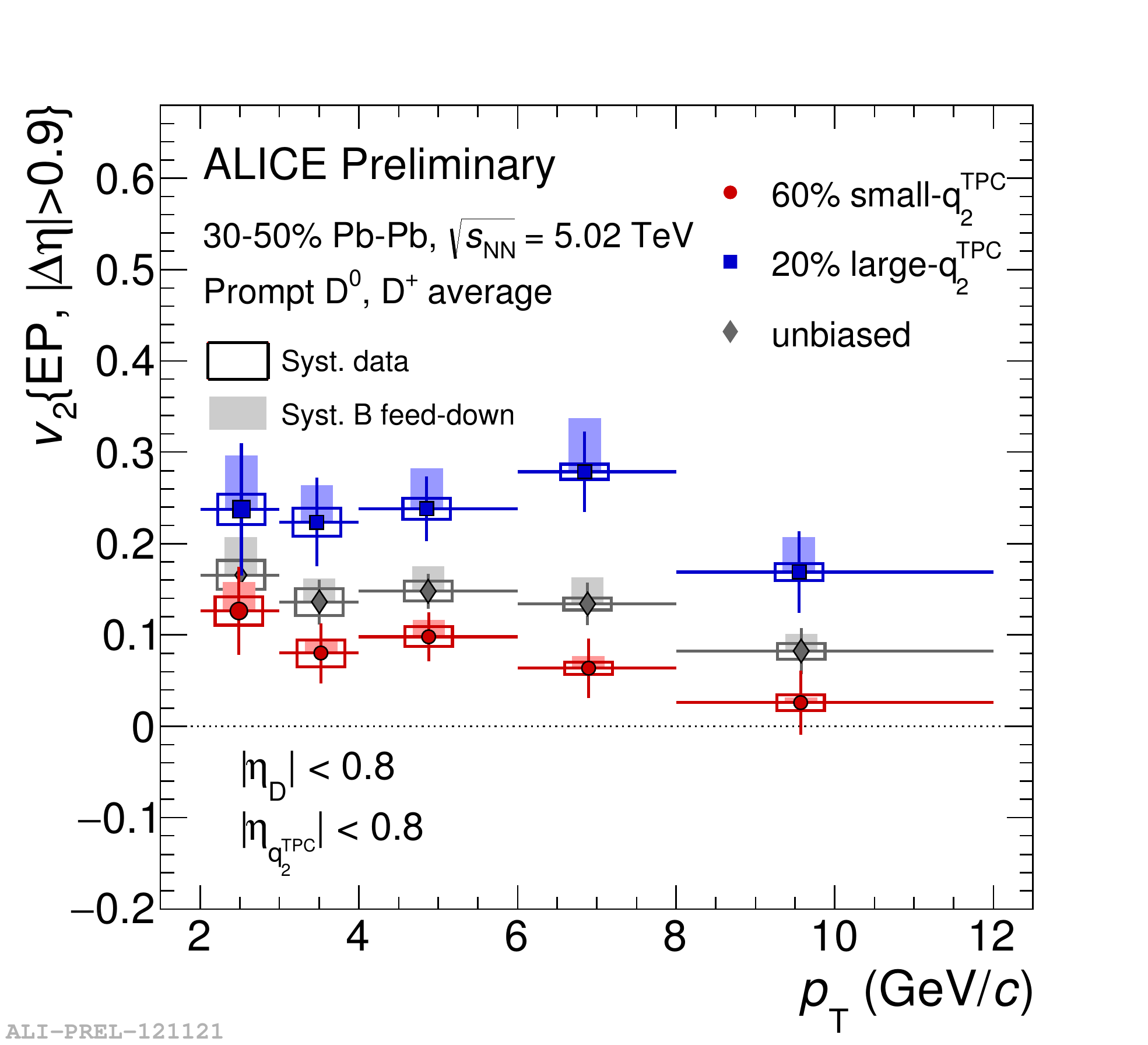}
\caption{\label{Dv2ESE}D meson $v_2$ in different classes of \qTPC.}
\end{minipage} 
\end{figure}

More differential quantities can be useful to further constrain models. In the case of the elliptic flow, events of the same centrality can produce a strong variation in the final state anisotropy. This can be quantified in by the reduced second--order flow vector $q_{2}^{\rm TPC}$. It grows with event multiplicity and flow strength \cite{Voloshin:2008dg}. As shown in figure \ref{qVector}, it fluctuates strongly within centrality classes. Selecting \qTPC as well as centrality classes of events gives a clear separation in the flow strength measured (Figure \ref{Dv2ESE}). Due to the fact that \qTPC is measured in at mid--rapidity, where the D mesons are measured as well, some correlations can be present. As a result, the measured $v_2$ decreases when an $\eta$ gap is introduced between the measurement of the D mesons and the \qTPC vector. This effect requires some additional careful analysis.

\begin{figure}[h]
\begin{minipage}{0.45\textwidth}
\vspace{1pc}
\includegraphics[height=13pc]{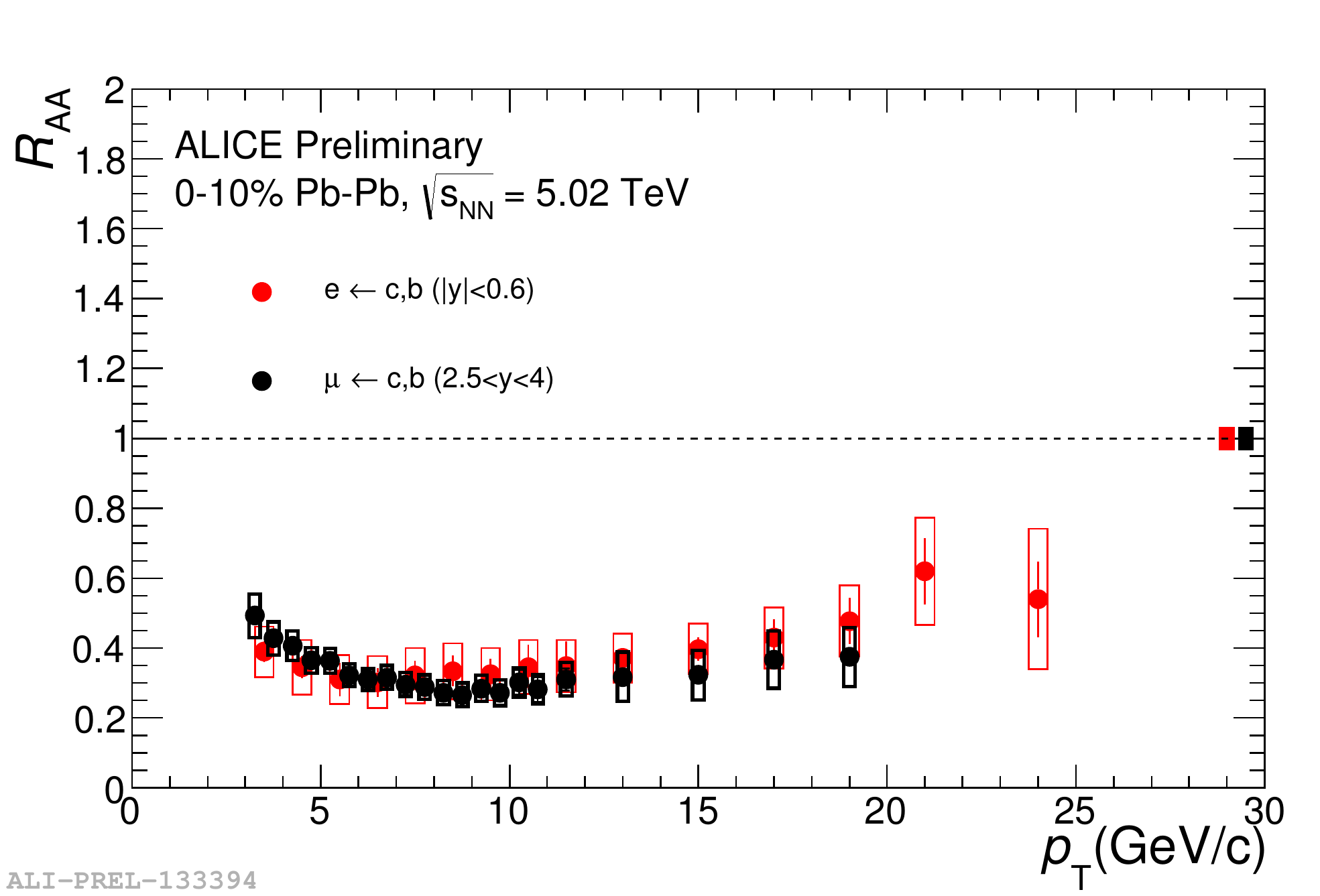}
\vspace{1pc}
\caption{\label{HFMRAA}\RAA of muons and electrons from heavy--flavour hadron decays at forward and mid--rapidity respectively.}
\end{minipage}\hspace{0.1\textwidth}%
\begin{minipage}{0.45\textwidth}
\includegraphics[height=15pc]{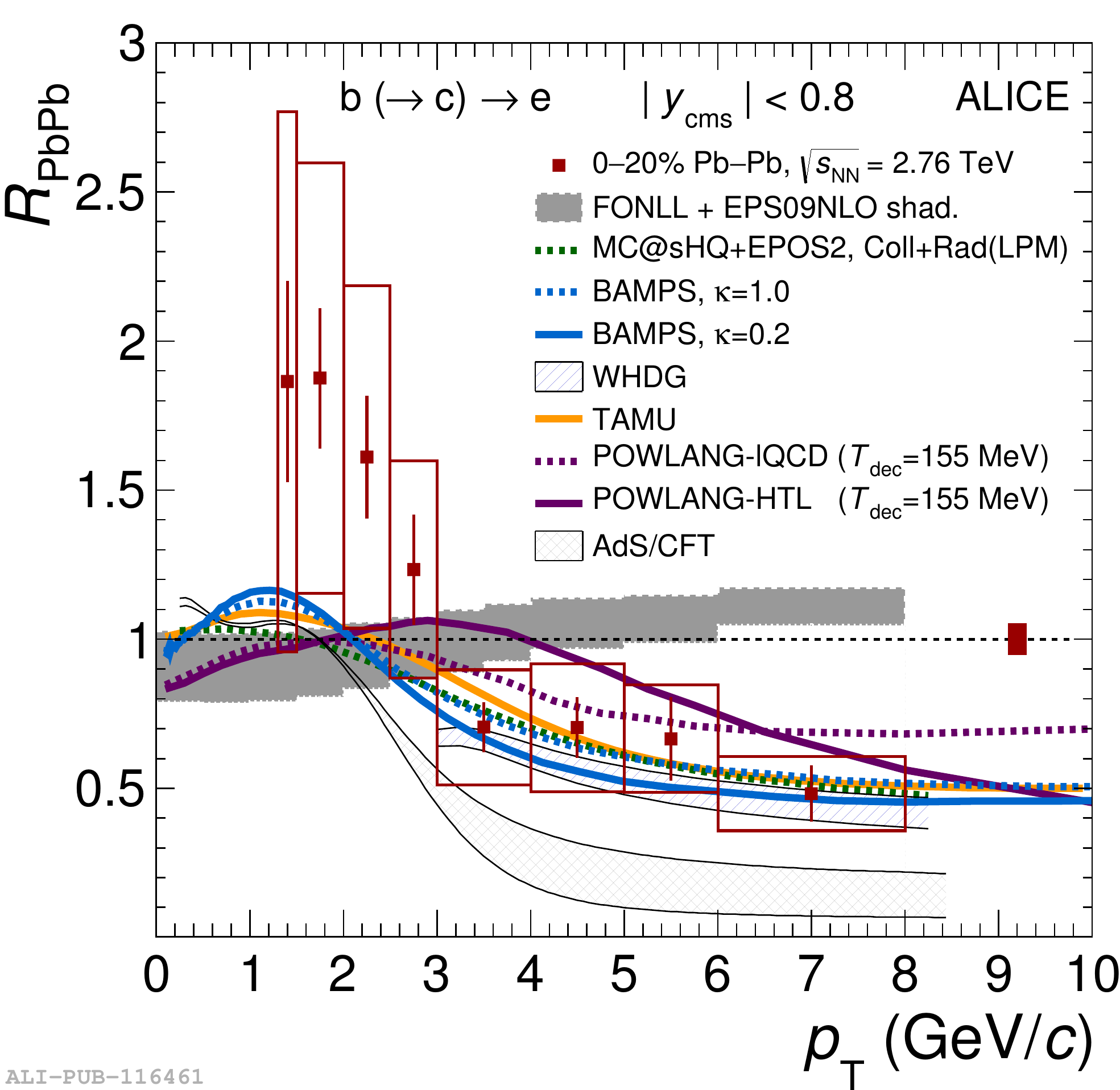}
\caption{\label{BtoeRAA}\RAA of electrons from heavy--flavour hadron decays \cite{Adam:2016wyz}.}
\end{minipage} 
\end{figure}

Even with the constraints from the nuclear modification factor and the elliptic flow, several diverse model calculations are able to describe the data reasonably. To complete the picture of heavy flavours in heavy ion collision, information about the mass dependence of the interaction is needed from the measurement of the beauty sector. Figure \ref{HFMRAA} shows the nuclear modification factor for electrons and muons from the decay of heavy--flavour hadrons. Towards higher \PT, the relative contribution of beauty increases, suggesting a suppression of beauty. This is further refined in the flavour--separated measurement shown in Figure \ref{BtoeRAA}. There is an indication for a suppression of electrons from beauty--hadron decays at the higher \PT~values measured, with the central points rising towards lower \PT. Several models can describe the measurement within uncertainties.

\section{Summary}

A comparison of the measurements of \PT--differential cross sections in pp and p--Pb collisions shows effects consistent with the expectations from cold nuclear matter effects, supporting the interpretation that the measured suppression and elliptic flow of heavy flavours in Pb--Pb collisions is a final state effect. This picture is complicated by the measurement of a significant elliptic flow for electrons from heavy--flavour decays even in p--Pb collisions. In addition, the measured production of charmed baryons in pp and p--Pb collisions is strongly underestimated by the model predictions. Future measurements may use more differential variables to constrain models further.

\section*{References}
\bibliography{ProceedingsBib}

\end{document}